\documentclass[twocolumn]{autart}    

\usepackage{graphicx}          

\usepackage{amssymb}
\usepackage{amsmath}
\usepackage{enumerate}
\usepackage{color}
\usepackage{algorithm}
\usepackage{algpseudocode}

\begin{document}

\begin{frontmatter}

\title{Data-Driven Robust Stabilization with Robust Domain of Attraction Estimate for Nonlinear Discrete-Time Systems \thanksref{footnoteinfo}} 

\thanks[footnoteinfo]{This paper was not presented at any IFAC
meeting. Corresponding author Y. Feng. Tel. +086-13567173448.}

\author[ZJUT]{Yongqiang Li}\ead{yqli@zjut.edu.cn}, 
\author[ZJUT]{Chaolun Lu}\ead{2111703050@zjut.edu.cn},    
\author[QDU]{Zhongsheng Hou}\ead{zshou@qdu.edu.cn},               
\author[ZJUT]{Yuanjing Feng}\ead{fyjing@zjut.edu.cn},

\address[ZJUT]{College of Information Engineering, Zhejiang University of Technology, Hangzhou, China} 
\address[QDU]{School of Automation, Qingdao University, Qingdao, China}


\begin{abstract}                          
	Nonlinear robust control is pursued by overcoming the drawback of linear robust control that it ignores available information about existing nonlinearities and the resulting controllers may be too conservative, especially when the nonlinearities are significant. However, most existing nonlinear robust control approaches just consider the affine nonlinear nominal model and thereby ignore available information about existing non-affine nonlinearities. When the general nonlinear nominal model is considered, the robust domain of attraction (RDOA) of closed-loops requires extensive investigation because it is hard to achieve the global stabilization. In this paper, we propose a new nonlinear robust control method based on Lyapunov function to stabilize a discrete-time uncertain system and to estimate the RDOA of closed-loops. First, a sufficient condition for robust stabilization of all plants in a plant set and estimation of the RDOA of all closed-loops is proposed. Then, to tackle the non-affine nonlinearities, a data-driven method of estimating the robust negative-definite domains (RNDD) is presented, and based on it the estimation of the RDOA of closed-loops and the resulting controller design are also given.
\end{abstract}

\end{frontmatter}

\section{Introduction}

Robust control theory is one of the most important branches of the modern control theory, due to its ability of dealing with the uncertainty describing how the "true" plant might differ from the nominal model. Most of robust control theory is linear (assume that the nominal model is linear) \cite{Safonov:2012_173,Petersen:2014_1315,Bhattacharyya:2017_45}. A disadvantage of linear robust control is that it ignores available information about existing nonlinearities, and the resulting controllers may be too conservative (especially when the nonlinearities are significant). A natural attempt to overcome this drawback is to allow the nominal model to be nonlinear and thereby pursue nonlinear robust control design \cite{Freeman:2008}.

Popular frameworks for robust nonlinear control include the Lyapunov min-max approach \cite{Corless:1993}, the nonlinear H$_\infty$ approach \cite{Basar:1995,Tsai:2018_3630}, the input-to-state stability approach \cite{Sontag:1995_351} and the robust backstepping approach \cite{Freeman:2008}. In these approaches, the nonlinear nominal model is affine with respect to the control, namely, affine nonlinear. Similar to the linear robust control, these approaches ignore available information about existing non-affine nonlinearities. A natural solution is to allow the nominal model to be non-affine nonlinear. 

For non-affine nonlinear system without uncertainty, due to the difficulty to achieve the global stabilization, the domain of attraction (DOA) of the closed-loop, an invariant set characterizing asymptotically stabilizable area around the equilibrium, requires extensive investigation \cite{Chen:2015_1314,Gering:2015_2231,Li:2014_79}. With the same reason, when the nominal model in robust control is non-affine nonlinear, the RDOA of the closed-loops also requires investigation under certain conditions. This issue is so difficult such that no result is published for robust controller design up to our knowledge. For system analysis, \cite{Goldsztejn:2019_371} proposes a method of estimating the RDOA for continuous-time autonomous systems under uncertainties. 

In this paper, we propose a new robust nonlinear control method based on Lyapunov function to stabilize a discrete-time uncertain system and estimate the RDOA of closed-loops. The uncertain system is described by a function set characterized by the nominal model and the corresponding modeling error bound. First, a sufficient condition for robust stabilization of all plants in the function set and estimation of the RDOA of all closed-loops is proposed. It is shown that if a state feedback controller belongs to the RNDD in the state-control space (RNDD-SC), in which all points make the difference of a given Lyapunov function to be negative-definite for all plants, it can asymptotically stabilize all plants. Meanwhile, any level-set of the Lyapunov function belonging to the RNDD in the state space (RNDD-S) can be an estimate of the RDOA of all closed-loops. Hence, if the RNDD-SC can be obtained, it is easy to find a robust controller and an estimate of the RDOA of closed-loops. However, due to nonlinearities, it is hard to obtain analytic solution of the RNDD-SC. Then, a data-driven method of estimating the RNDD-SC is proposed. The idea is much simple. The state-control space is partitioned into disjoint cells. The estimate of the RNDD-SC consists of cells in which all data points satisfying specific conditions. We would like stress that the present paper considers a given Lyapunov function and addresses the problem of robust stabilization and estimation of RDOA from it. The problem of finding good Lyapunov functions is not in the scope of this paper.

The rest of this paper is organized as follows. In Section~\ref{sec:problem}, the control problem is formulated. In Section~\ref{sec:sufficinet condition}, sufficient conditions for robust stabilization and estimation of the RDOA of all closed-loops are proposed. In Section~\ref{sec:stabilization}, a data-driven robust stabilization with the RDOA estimation is derived. Finally, the conclusion is drawn in Section~\ref{sec:conclusion}.

\textbf{Notation: } For $x \in \mathbb{R}^n$ and $u \in \mathbb{R}^m$, $(x;u)$ represents a new vector in $\mathbb{R}^{n+m}$. For $x_1, x_2 \in \mathbb{R}^n$, $x_1 \leq x_2$ means $x_1$ is less than or equal to $x_2$ element by element.

\section{Problem formulation} \label{sec:problem}

Consider the plant set
\begin{eqnarray}
	&& \!\! \mathfrak{F} = \Big\{ f: \mathbb{R}^n \!\times\! \mathbb{R}^m \!\to \!	\mathbb{R}^n \Big| f(0,0) = 0, \nonumber \\ 
	&& \quad \hat{f}(x,u) - \delta(x,u) \leq f(x,u) \leq \hat{f}(x,u) + \delta(x,u) \Big\}, \label{eq:plant_set}
\end{eqnarray}
where nominal model $\hat{f}: \mathbb{R}^n \times \mathbb{R}^m \to \mathbb{R}^n$ and modeling error bound $\delta: \mathbb{R}^n \times \mathbb{R}^m \to \mathbb{R}^n_+$ are continuous, satisfying $\hat{f}(0,0) = 0$ and $\delta(0,0) = 0$, $x \in \mathbb{R}^n$ is state and $u \in \mathbb{R}^m$ is control input. The control objective is to find a robust controller $\mu$ and an estimate of the RDOA such that, $\forall f \in \mathfrak{F}$, the closed-loop $x(k+1) = f(x(k),\mu(x(k)))$ is asymptotically stable at the origin for all initial state in the estimate of the RDOA.

\begin{rem}
	The nominal model error bound is described by a function rather than a constant. This can take advantage of available information about existing nonlinearities of the nominal model error bound and avoid that the resulting controllers and estimates of the RDOA may be too conservative. There are data-driven modeling methods which can give such error bound, \textit{e.g.}, Gaussian processes regression \cite{Rasmussen:2006}.
\end{rem}

\section{Sufficient condition for robust stabilization with robust DOA estimate} \label{sec:sufficinet condition}

\begin{figure*}
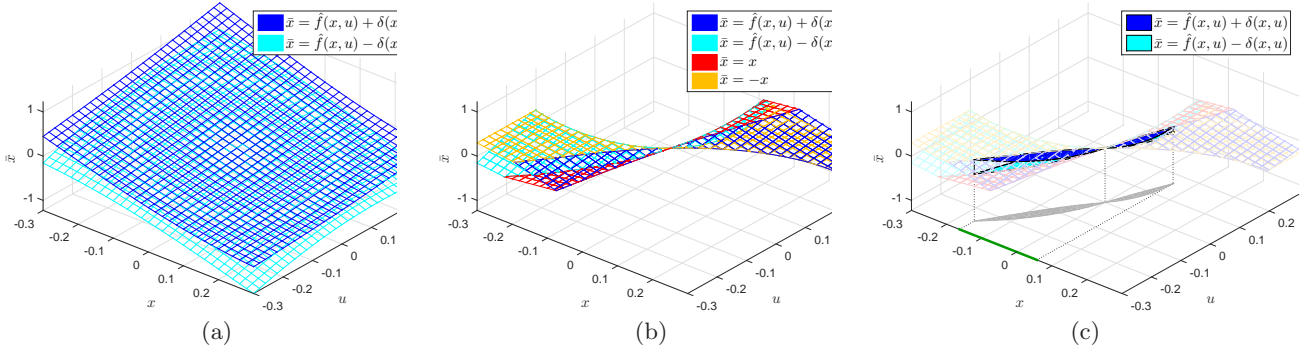

	\begin{center}
		\includegraphics[width=0.32\textwidth]{illustrateTheorem_F.eps}
		\includegraphics[width=0.32\textwidth]{illustrateTheorem_F_pL.eps}
		\includegraphics[width=0.32\textwidth]{illustrateTheorem_F_allpL.eps}\\
		\parbox[c]{0.32\textwidth}{\footnotesize \centering (a)}
		\parbox[c]{0.32\textwidth}{\footnotesize \centering (b)}
		\parbox[c]{0.32\textwidth}{\footnotesize \centering (c)}
		\caption{(a) Plant set $\Pi_{\mathfrak{F}}$. (b) Negative-definite domain $\tilde{\Pi}_{\mathfrak{F}} (L)$. (c) Robust negative-definite domains $\Pi_{\mathfrak{F}} (L)$, $\mathbb{W}_{\mathfrak{F}} (L)$ and $\mathbb{X}_{\mathfrak{F}} (L)$.}
		\label{fig:illustrateTheorem}
	\end{center}
\end{figure*}

For given nominal model $\hat{f}$ and modeling error bound $\delta$, the plant set $\mathfrak{F}$ defined in \eqref{eq:plant_set} can be alternatively described as a domain 
\begin{eqnarray}
	 \Pi_{\mathfrak{F}} \!=\! \Big\{ (\bar{x};x;u)\Big| \hat{f}(x,\!u) \!-\! \delta(x,\!u) \!\leq\! \bar{x} \!\leq\! \hat{f}(x,\!u) \!+\! \delta(x,\!u) \Big\} \label{eq:Pi}
\end{eqnarray}
in $(2n+m)$-dimensional space, where $\bar{x} \in \mathbb{R}^n$ denotes the future state (\textit{i.e.} the state at the next time step), $x \in \mathbb{R}^n$ and $u \in \mathbb{R}^m$ denote the current state and control input. Considering a continuous positive-definite function $L: \mathbb{R}^n \to \mathbb{R}$, we define the negative-definite domain in the state-state-control space (NDD-SSC) $\tilde{\Pi}_{\mathfrak{F}}(L)$ as
\begin{eqnarray}
	\tilde{\Pi}_{\mathfrak{F}}(L) = \Big\{ (\bar{x};x;u) \in \Pi_{\mathfrak{F}} \Big| L(\bar{x}) - L(x) < 0 \Big\}. \label{eq:nrNDD_Pi}
\end{eqnarray} 
Although any point $(\bar{x};x;u) \in \tilde{\Pi}_{\mathfrak{F}}(L)$ makes the time difference of $L$ to be negative-definite (\textit{i.e.}, $L(\bar{x}) - L(x) < 0$), $\tilde{\Pi}_{\mathfrak{F}}(L)$ is not robust for plant set \eqref{eq:plant_set}. Because it is not guaranteed that, for given $(x;u)$, $\forall f \in \mathfrak{F}$ satisfies $L(f(x,u)) - L(x) < 0$. In order to define the robust NDD-SSC (RNDD-SSC), we first define the future state set $\bar{\mathbb{X}}_{\mathfrak{F}} (x,u)$ for the given $(x;u)$ as follows.
\begin{eqnarray}
	\bar{\mathbb{X}}_{\mathfrak{F}} (x,\!u) \!= \!\Big\{\bar{x} \!\in\! \mathbb{R}^n \Big| \hat{f}(\!x,\!u\!) \!-\! \delta(\!x,\!u\!) \!\leq\! \bar{x} \!\leq\! \hat{f}(\!x,\!u\!) \!+\! \delta(\!x,\!u\!) \!\Big\}. \label{eq:X_bar(x,u)}
\end{eqnarray}
With $\bar{\mathbb{X}}_{\mathfrak{F}} (x,u)$, the RNDD-SSC $\Pi_{\mathfrak{F}}(L)$ is defined as
\begin{eqnarray}
	&& \!\! \Pi_{\mathfrak{F}}(L) = \Big\{ (\bar{x};x;u) \in \tilde{\Pi}_{\mathfrak{F}} (L) \Big| \nonumber \\
	&& \quad  \forall \bar{x}^{\prime} \in \bar{\mathbb{X}}_{\mathfrak{F}} (x,u), L(\bar{x}^{\prime}) - L(x) < 0, \bar{x} \in \bar{\mathbb{X}}_{\mathfrak{F}} (x,u) \Big\}. \label{eq:NDD_Pi}
\end{eqnarray}
Projecting $\Pi_{\mathfrak{F}}(L)$ along the future state space onto the state-control space, the RNDD in the state-control space (RNDD-SC) $\mathbb{W}_{\mathfrak{F}} (L)$ is defined as
\begin{eqnarray}
	\mathbb{W}\!_{\mathfrak{F}}\! (L) \!=\! \Big\{\!(x;\!u) \!\in\! \mathbb{R}^{n+m} \Big| \forall \bar{x} \!\in\! \bar{\mathbb{X}}_{\mathfrak{F}} (x,\!u), L\big(\!\bar{x}\!\big) \!-\! L(\!x\!) \!<\! 0 \Big\}.\! \label{eq:NDD_W} 
\end{eqnarray}
The RNDD-SC $\mathbb{W}_{\mathfrak{F}} (L)$ is an open set, because its boundary $\{(x;u) | \forall \bar{x} \in \bar{\mathbb{X}}_{\mathfrak{F}} (x,u), L(\bar{x}) - L(x) \leq 0 \} \backslash \mathbb{W}_{\mathfrak{F}} (L)$ is not its subset, where $A \backslash B = \{a \in A | a \notin B \}$. It is obvious that the origin $(0;0) \in \mathbb{R}^{n+m}$ is in the boundary of $\mathbb{W}_{\mathfrak{F}} (L)$. Due to $\bar{\mathbb{X}}_{\mathfrak{F}} (0,0) = \{0\}$ and the continuity of $\hat{f}, \delta$ and $L$, there is a subset of the neighborhood of the origin $(0;0)$ contained by $\mathbb{W}_{\mathfrak{F}} (L)$ although the origin $(0;0)$ is not contained by $\mathbb{W}_{\mathfrak{F}} (L)$. Projecting $\mathbb{W}_{\mathfrak{F}}(L)$ along the control space onto the state space, the RNDD in the state space (RNDD-S) $\mathbb{X}_{\mathfrak{F}} (L)$ is defined as
\begin{eqnarray}
\mathbb{X}_{\mathfrak{F}} (L) = \Big\{x \in \mathbb{R}^{n} \Big| \exists u \in \mathbb{R}^m, (x;u) \in \mathbb{W}_{\mathfrak{F}}(L) \Big\}. \label{eq:NDD_X} 
\end{eqnarray}
The RNDD-S $\mathbb{X}_{\mathfrak{F}} (L)$ does not contain the origin $0 \in \mathbb{R}^n$ but contains a neighborhood of the origin. Examples of plant set $\Pi_{\mathfrak{F}}$, NDD-SSC $\tilde{\Pi}_{\mathfrak{F}} (L)$, RNDD-SSC $\Pi_{\mathfrak{F}} (L)$, RNDD-SC $\mathbb{W}_{\mathfrak{F}} (L)$ and RNDD-S $\mathbb{X}_{\mathfrak{F}} (L)$ are given in Example~\ref{exmp:illuminate}.

\begin{exmp} \label{exmp:illuminate}
	Consider the nominal model $\hat{f}(x,u) = -\sin(2x) - xu - 0.2x - u^2 + u$ and the model error bound $\delta = 1 - \exp \left(-2(x^2 + u^2)\right)$, where $x \in \mathbb{R}$ and $u \in \mathbb{R}$.The plant set $\Pi_{\mathfrak{F}} \subset \mathbb{R}^3$ defined in \eqref{eq:Pi} is shown in Figure~\ref{fig:illustrateTheorem} (a). Considering the positive-definite function $L(x) = x^2$, the NDD-SSC $\tilde{\Pi}_{\mathfrak{F}}(L) \subset \mathbb{R}^3$ defined in \eqref{eq:nrNDD_Pi} is shown in Figure~\ref{fig:illustrateTheorem} (b). Any $(\bar{x};x;u) \in \mathbb{R}^3$ between the surfaces $\bar{x} = x$ and $\bar{x} = -x$ satisfies $\bar{x}^2 - x^2 < 0$. Hence, boundaries of $\tilde{\Pi}_{\mathfrak{F}} (L) \subset \mathbb{R}^3$ consist of four surfaces: $\bar{x} = \hat{f}(x,u) + \delta(x,u), \bar{x} = \hat{f}(x,u) - \delta(x,u), \bar{x} = x$ and $\bar{x} = -x$.
	
	The RNDD-SSC $\Pi_{\mathfrak{F}}(L) \subset \mathbb{R}^3$ defined in \eqref{eq:NDD_Pi} is shown in Figure~\ref{fig:illustrateTheorem} (c) denoted by the domain between the blue patch in the surface $\bar{x} = \hat{f}(x,u) + \delta(x,u)$ and the cyan patch in the surface $\bar{x} = \hat{f}(x,u) - \delta(x,u)$. It should be noted that the boundary of the blue patch is identical with the one of the cyan patch in the state-control space. The RNDD-SC $\mathbb{W}_{\mathfrak{F}}(L) \subset \mathbb{R}^2$ defined in \eqref{eq:NDD_W} is shown in Figure~\ref{fig:illustrateTheorem}(c) denoted by the gray region in $(x$-$u)$-plane. The RNDD-S $\mathbb{X}_{\mathfrak{F}} (L) \in \mathbb{R}$ is shown in Figure~\ref{fig:illustrateTheorem} (c) denoted by the green line segment in the $x$-axis.
\end{exmp}

From \eqref{eq:NDD_W}, it is straightforward that, if $(x(k);u(k)) \in \mathbb{W}_{\mathfrak{F}}(L)$, then, $\forall f \in \mathfrak{F}, L(f(x(k),u(k))) - L(x(k)) < 0$. Based on this and \eqref{eq:NDD_X}, one may concludes that closed-loops of all plants in $\mathfrak{F}$ are asymptotically stable for any initial state in $\mathbb{X}_{\mathfrak{F}} (L)$ if the state feedback controller belongs to $\mathbb{W}_{\mathfrak{F}}(L)$. Unfortunately, this conclusion is wrong. Because it can not be guaranteed that the future state is still in $\mathbb{X}_{\mathfrak{F}}(L)$ at the next time step. Once, the future state is outside of $\mathbb{X}_{\mathfrak{F}}(L)$, the condition that the difference of $L$ is negative-definite is no longer satisfied. To solve this problem, it is needed to find an invariant subset of $\mathbb{X}_{\mathfrak{F}}(L)$ as the estimate of closed-loops' DOA. Level-set $\mathbb{X}_{\mathrm{ls}}(L,\alpha)$ of positive-definite function $L$ with constant $\alpha > 0$ is just an invariant set, which has the property that if the current state is in $\mathbb{X}_{\mathrm{ls}}(L,\alpha)$, then the next time step state is also in $\mathbb{X}_{\mathrm{ls}}(L,\alpha)$. Hence, any level-set $\mathbb{X}_{\mathrm{ls}}(L,\alpha) \subset \mathbb{X}_{\mathfrak{F}}(L)$ can be an estimate of closed-loops' DOA. This idea is summarized in Theorem~\ref{thm:robust_stab}.

\begin{thm} \label{thm:robust_stab}
	For plant set \eqref{eq:plant_set}, if a positive-definite function $L: \mathbb{R}^n \to \mathbb{R}$, a constant $\alpha \in \mathbb{R}_+$ and a state feedback controller $\mu: \mathbb{R}^n \to \mathbb{R}^m$ exist such that 
	\begin{eqnarray}
		&&\!\! \mu(0) = 0, (x;\mu(x)) \in \mathbb{W}_{\mathfrak{F}} (L), \forall x \in \mathbb{X}_{\mathfrak{F}} (L),  \label{eq:thm:robust_stab:mu} \\
		&&\!\! \mathbb{X}_{\mathrm{ls}} (L,\alpha) = \Big\{x \in \mathbb{R}^n \ \Big| \ L(x) \leq \alpha \Big\} \subset \mathbb{X}_{\mathfrak{F}} (L) \cup \{0\}, \label{eq:thm:robust_stab:X_L_alpha}
	\end{eqnarray}
	then, $\forall f \in \mathfrak{F}$, closed-loop $x(k+1) = f(x(k),\mu(x(k)))$ is asymptotically stable for any initial state in $\mathbb{X}_{\mathrm{ls}} (L,\alpha)$, where $\mathbb{W}_{\mathfrak{F}} (L)$ and $\mathbb{X}_{\mathfrak{F}} (L)$ are defined in \eqref{eq:NDD_W} and \eqref{eq:NDD_X}.
\end{thm}

\begin{pf}
	From \eqref{eq:NDD_W}-\eqref{eq:thm:robust_stab:X_L_alpha}, it follows that
	\begin{eqnarray}
	\forall \!f\! \!\in\! \mathfrak{F}, \forall x \!\in\! \mathbb{X}_{\text{ls}} (L,\alpha) \backslash \{0\}, L\big(f\!\left(x,\mu(x)\right)\!\big) \!-\! L(x) \!<\! 0. \label{eq:thm:pf:negative-definite in X_ls}
	\end{eqnarray}
	For any $f \in \mathfrak{F}$, let $\phi(x_0,k)$ denote the solution of $x(k+1) = f\big(x(k),\mu(x(k))\big)$ at time $k$ with the initial state $x_0$. From \eqref{eq:thm:pf:negative-definite in X_ls}, it follows that, $\forall x_0 \in \mathbb{X}_{\text{ls}} (L,\alpha) \backslash \{0\}$,
	\begin{eqnarray}
	L(\phi(x_0,k+1)) < L(\phi(x_0,k)) \leq L(x_0) \leq \alpha. \label{eq:thm:pf:L(k+1)<L(k)}
	\end{eqnarray}
	The above relation shows that $\phi(x_0,k)$ starting in $\mathbb{X}_{\text{ls}} (L,\alpha)$ remains in $\mathbb{X}_{\text{ls}} (L,\alpha)$, namely, $\mathbb{X}_{\text{ls}} (L,\alpha)$ is an invariant set of $x(k+1) = f\big(x(k),\mu(x(k))\big), \forall f \in \mathfrak{F}$. Because $\mathbb{X}_{\text{ls}} (L,\alpha)$ is invariant, \eqref{eq:thm:pf:L(k+1)<L(k)} also shows that, $\forall x_0 \in \mathbb{X}_{\text{ls}} (L,\alpha) \backslash \{0\}$, $L(\phi(x_0,k))$ is monotonically decreasing with time. And because $L$ is positive-definite, $L(\phi(x_0,k))$ is bounded from below by zero. Hence, $\forall f \in \mathfrak{F}, \forall x_0 \in \mathbb{X}_{\text{ls}} (L,\alpha), \lim_{k \to \infty} L(\phi(x_0,k)) = 0$. This means that $\forall f \in \mathfrak{F}, \forall x_0 \in \mathbb{X}_{\text{ls}} (L,\alpha), \lim_{k \to \infty} \phi(x_0,k) = 0$ (this can be proven by reductio ad absurdum. For details, see the proof of Theorem 13.2 in \cite{Haddad:2008}). \hfill $\blacksquare$
\end{pf}

\begin{rem}
	It should be noted that the result of Theorem~\ref{thm:robust_stab} is conservative because, for the given $(x;u)$, $L(\bar{x}) - L(x) < 0$ must satisfies for all $\bar{x} \in \bar{\mathbb{X}}_{\mathfrak{F}}(x,u)$, namely, $L(f(x,u)) - L(x) < 0$ must satisfies for all $f \in \mathfrak{F}$. One way to alleviate this problem is selecting a good Lyapunov function. For different Lyapunov functions, RNDDs are totally different. Enlargement of RNDDs can be achieved by selecting a good Lyapunov function from a positive-definite function set, \textit{e.g.}, sum-of-square polynomials. This is our future work.
\end{rem}

\section{Data-driven robust asymptotic stabilization with robust DOA estimate} \label{sec:stabilization}

From Theorem~\ref{thm:robust_stab}, for a given positive-definite function $L: \mathbb{R}^n \to \mathbb{R}$, if the RNDD-SC $\mathbb{W}_{\mathfrak{F}}(L) \subset \mathbb{R}^n \times \mathbb{R}^m$ and RNDD-S $\mathbb{X}_{\mathfrak{F}}(L) \subset \mathbb{R}^n$ are obtained, it is easy to find a robust controller and an estimate of the RDOA of closed-loops. However, due to nonlinearities of $\hat{f}$, $\delta$ and $L$, it is hard to obtain analytic solutions of the RNDDs. In this section, first, a data-driven method of estimating the RNDDs is proposed. Then, based on the estimates of RNNDs, methods of estimating the RDOA and finding the robust controller are also introduced.

\subsection{Robust NDDs estimation}

\begin{figure}
	\begin{center}
		\includegraphics[width=0.4\textwidth]{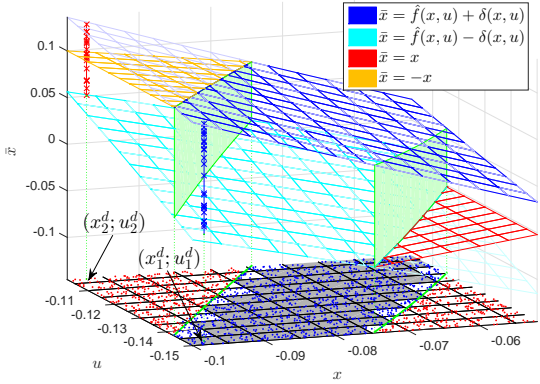}
		\caption{Illumination of estimating robust NDD $\hat{\mathbb{W}}_{\mathfrak{F}}(L)$} \label{fig:illumination_W_hat}
	\end{center}
\end{figure}

The idea of estimating the RNDD-SC $\mathbb{W}_{\mathfrak{F}}(L)$ is simple. First, generate the sample data set $W^d$ of the interested region $\mathbb{W} \subset \mathbb{R}^{n+m}$ under the uniform distribution. Then, find the sample data set $W^d_{\mathfrak{F}} (L) \subset W^d$ of $\mathbb{W}_{\mathfrak{F}}(L)$. Finally, The $\mathbb{W}$ is partitioned into disjoint cells and the estimate $\hat{\mathbb{W}}_{\mathfrak{F}}(L)$ of $\mathbb{W}_{\mathfrak{F}}(L)$ consists of cells in which all data points belong to $W^d_{\mathfrak{F}} (L)$.

First, let hyper-rectangle $\mathbb{W} \subset \mathbb{R}^{n+m}$ denote the interested region. A sample data set $W^d$ of $\mathbb{W}$ can be generated, in which each data point $(x^d;u^d) \in \mathbb{R}^{n+m}$ is drawn from the uniform distribution on $\mathbb{W}$ and the number of data points is $N_{\text{xu}}$. An example of $W^d$ is shown in Figure~\ref{fig:illumination_W_hat} denoted by blue and red dots in $(x$-$u)$-plane ($\hat{f}, \delta$ and $L$ are the same as those in Example~\ref{exmp:illuminate}).

Then, we aim to find the sample data set $W^d_{\mathfrak{F}} (L) \subset W^d$ of $\mathbb{W}_{\mathfrak{F}}(L)$. From \eqref{eq:NDD_W}, any $(x;u) \in \mathbb{W}_{\mathfrak{F}}(L)$ satisfies, $\forall \bar{x} \in \bar{\mathbb{X}}_{\mathfrak{F}}(x,u)$, $L(\bar{x}) - L(x) < 0$. Based on this, the idea of verifying whether a data point $(x^d; u^d) \in W^d$ belongs to $\mathbb{W}_{\mathfrak{F}} (L)$ is as following. Firstly, for each $(x^d; u^d) \in W^d$, a sample data set $\bar{X}_{\mathfrak{F}}^d(x^d,u^d)$ of $\bar{\mathbb{X}}_\mathfrak{F} (x^d,u^d)$ is generated, in which each data point $\bar{x}^d \in \mathbb{R}^{n}$ is drawn from the uniform distribution on $\bar{\mathbb{X}}_\mathfrak{F} (x^d,u^d)$ and the number of data points is $N_{\mathrm{\bar{x}}}$. Secondly, for each $(x^d; u^d) \in W^d$, if $L(\bar{x}^d) - L(x^d) < 0, \forall \bar{x}^d \in \bar{X}_{\mathfrak{F}}^d(x^d,u^d)$, $(x^d; u^d)$ is deemed to belong to $\mathbb{W}_{\mathfrak{F}} (L) $. Hence, the sample data set
\begin{eqnarray}
&&\!\! W^d_{\mathfrak{F}} (L) = \Big\{(x^d;u^d) \in W^d \Big| \forall \bar{x}^d \in \bar{X}_{\mathfrak{F}}^d(x^d,u^d), \nonumber \\
&& \qquad L(\bar{x}^d) - L(x^d) < 0 \Big\} \label{eq:W^d_F(L)}
\end{eqnarray}
of $\mathbb{W}_{\mathfrak{F}} (L)$ can be obtained. The example of $W^d_{\mathfrak{F}} (L)$ is shown in Figure~\ref{fig:illumination_W_hat} denoted by blue dots in $(x$-$u)$-plane. For data point $(x^d_1;u^d_1) \in \mathbb{R}^2$, $\bar{\mathbb{X}}_{\mathfrak{F}}(x^d_1,u^d_1)$ is denoted by the blue line segment perpendicular to $(x$-$u)$-plane and sample data set $\bar{X}_{\mathfrak{F}}^d(x^d_1,u^d_1)$ is denoted by blue 'x'. For $(x^d_2;u^d_2) \in \mathbb{R}^2$, $\bar{\mathbb{X}}_{\mathfrak{F}}(x^d_2,u^d_2)$ is denoted by the red line segment and sample data set $\bar{X}_{\mathfrak{F}}^d(x^d_2,u^d_2)$ is denoted by red 'x'. Because all $\bar{x}^d \in \bar{X}_{\mathfrak{F}}^d(x^d_1,u^d_1)$ are between the surfaces $\bar{x} = x$ and $\bar{x} = -x$, namely $L(\bar{x}^d) - L(x^d) < 0$, $(x^d_1;u^d_1)$ is collected by data set $W^d_{\mathfrak{F}} (L)$, while $(x^d_2;u^d_2)$ is not.

Finally, the interested region $\mathbb{W}$ is partitioned into disjoint cells. Here, we apply a uniform grid over $\mathbb{W}$ and each cell is a rectangle or hyper-rectangle. Suppose there are $N_\text{c}$ cells in the grid and each cell is denoted by $\mathbb{C}_i, i = 1,2,\cdots, N_{\text{c}}$. An inner approximation $\hat{\mathbb{W}}_{\mathfrak{F}} (L)$ of $\mathbb{W}_{\mathfrak{F}} (L)$ can be obtained by combining all cells only containing data points in $W^d_{\mathfrak{F}} (L)$. The example of $\hat{\mathbb{W}}_{\mathfrak{F}} (L)$ is shown in Figure~\ref{fig:illumination_W_hat} denoted by gray rectangles in $(x$-$u)$-plane. Figure~\ref{fig:illumination_W_hat} also shows $\mathbb{W}_{\mathfrak{F}} (L)$ denoted by the domain between the two green curves in $(x$-$u)$-plane and $\Pi_{\mathfrak{F}} (L)$ denoted by the 3-dimensional domain between the blue surface, the cyan surface and the two green surfaces.

The above procedure is summarized in Algorithm~\ref{alg:est_W_F(L)}.

\begin{algorithm} 
	\caption{Robust NDDs estimation algorithm} \label{alg:est_W_F(L)}
	For the given positive-definite function $L$, plant set $\mathfrak{F}$ and grid over $\mathbb{W}$, estimates $\hat{\mathbb{W}}_{\mathfrak{F}} (L)$ and $\hat{\mathbb{X}}_{\mathfrak{F}} (L)$ of RNDDs can be obtained as the following:
	
	\begin{algorithmic}[1]
		\State Generate the sample data set $W^d$ whose data points are drawn from the uniform distribution on $\mathbb{W}$;
		
		\State For each data point $(x^d;u^d) \in W^d$, generate the sample data set $\bar{X}^d_{\mathfrak{F}}(x^d,u^d)$ whose data points are drawn from the uniform distribution on $\bar{\mathbb{X}}_{\mathfrak{F}}(x^d,u^d)$;
		
		\State Find the sample data set $W^d_{\mathfrak{F}}(L)$ defined in \eqref{eq:W^d_F(L)} by selecting data point $(x^d;u^d) \in W^d$ which satisfies $L(\bar{x}^d) - L(x^d) < 0, \forall \bar{x}^d \in \bar{X}_{\mathfrak{F}}^d(x^d,u^d)$;
		
		\State Obtain the estimate $\hat{\mathbb{W}}_{\mathfrak{F}} (L)$ of $\mathbb{W}_{\mathfrak{F}} (L) \subset \mathbb{W}$ by combining all cells only containing data points in $W^d_{\mathfrak{F}} (L)$;
		
		\State Obtain the estimate $\hat{\mathbb{X}}_{\mathfrak{F}} (L)$ of $\mathbb{X}_{\mathfrak{F}} (L)$ by projecting $\hat{\mathbb{W}}_{\mathfrak{F}} (L)$ along the control space onto the state space.
	\end{algorithmic}
\end{algorithm}

\begin{rem}
	In Step 1 and 2, Algorithm~\ref{alg:est_W_F(L)} generates $N_{xu}(n+m)+N_{xu}N_{\bar{x}}n$ random numbers. In Step 3, the number of verification of $L(\bar{x})^d - L(x^d) < 0$ is $N_{xu}N_{\bar{x}}$. In Step 4, the number of verification whether or not a cell satisfies that all data points in it belong to $W^d_{\mathfrak{F}} (L)$ is $N_c$. It is hard to find a quantitative analysis result about the precision of Algorithm~\ref{alg:est_W_F(L)} because it uses the random sampling and griding method to approximate the desired domain. We can only declare that, in order to obtain a good estimation, $N_{xu}, N_{\bar{x}}$ and $N_c$ must be enough large.
\end{rem}

\subsection{Estimating robust DOA of closed-loops and designing controller}

The RNDD-SC $\mathbb{W}_{\mathfrak{F}} (L)$ defined in \eqref{eq:NDD_W} is an open set and the origin $(0;0) \in \mathbb{R}^{n+m}$ is in the boundary of $\mathbb{W}_{\mathfrak{F}} (L)$. There is no cell belonging to the inner approximation $\hat{\mathbb{W}}_{\mathfrak{F}} (L)$ nearby the origin $(0;0) \in \mathbb{R}^{n+m}$ and there is a small neighborhood $\mathbb{X}_0 \subset \mathbb{R}^n$ of the origin in the state space that is not contained by the projection $\hat{\mathbb{X}}_{\mathfrak{F}} (L)$ of $\hat{\mathbb{W}}_{\mathfrak{F}} (L)$. Hence, the condition \eqref{eq:thm:robust_stab:X_L_alpha} of estimating RDOA in Theorem~\ref{thm:robust_stab} is modified as $\mathbb{X}_{\text{ls}} (L,\alpha) \subset \hat{\mathbb{X}}_{\mathfrak{F}} (L) \cup \mathbb{X}_0$. Note that the volume of $\mathbb{X}_{\text{ls}} (L,\alpha)$ is increasing as $\alpha$ is increasing for a given Lyapunov function $L$. Hence, the largest estimate of RDOA can be obtained by solving the optimization problem
\begin{eqnarray}
	\max_{\alpha \in \mathbb{R}_+} \alpha \quad \text{subject to } \mathbb{X}_{\text{ls}} (L,\alpha) \subset \hat{\mathbb{X}}_{\mathfrak{F}} (L) \cup \mathbb{X}_0.  \label{eq:optimization_alpha}
\end{eqnarray}  
In order to verify the constrains in \eqref{eq:optimization_alpha}, it is needed to
estimate the level-set $\mathbb{X}_{\text{ls}} (L,\alpha)$. With the same idea of estimating the RNDD-SC $\mathbb{W}_{\mathfrak{F}} (L)$, the estimate of  $\mathbb{X}_{\text{ls}} (L,\alpha)$ can be obtained. Let $\alpha^\ast$ be the solution of \eqref{eq:optimization_alpha}. The estimate of RDOA is $\mathbb{X}_{\text{ls}} (L,\alpha^\ast)$.

Replacing the RNDD-SC $\mathbb{W}_{\mathfrak{F}} (L) \subset \mathbb{R}^{n+m}$ with its estimate $\hat{\mathbb{W}}_{\mathfrak{F}} (L)$, from Theorem~\ref{thm:robust_stab}, we know that any controller $\mu$ belonging to $\hat{\mathbb{W}}_{\mathfrak{F}} (L)$ can stabilize all plants in $\mathfrak{F}$. A simple way to find a controller $\mu$ belonging to $\hat{\mathbb{W}}_{\mathfrak{F}} (L)$ is that, first, select a controller training set belonging to $\hat{\mathbb{W}}_{\mathfrak{F}} (L)$; then, obtain the controller $\mu$ with a function estimation method, such as interpolation, Gaussian processes regression and so on. When the trend of the training data points is smooth enough and $\mu(0) = 0$ is constrained, it can be guaranteed that the controller obtained from the function estimator belongs to $\hat{\mathbb{W}}_{\mathfrak{F}} (L)$.

\begin{figure}
	\begin{center}
		\includegraphics[width=0.23\textwidth]{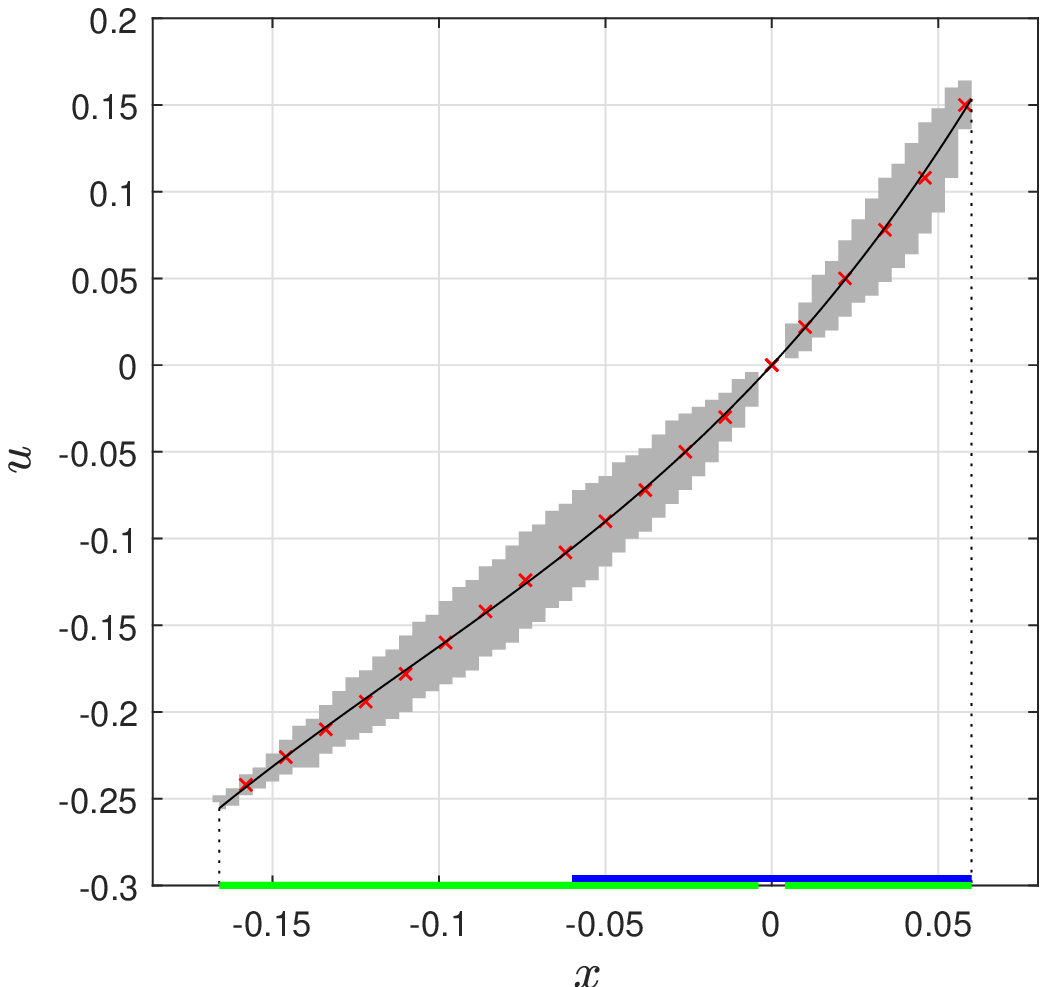}
		\includegraphics[width=0.23\textwidth]{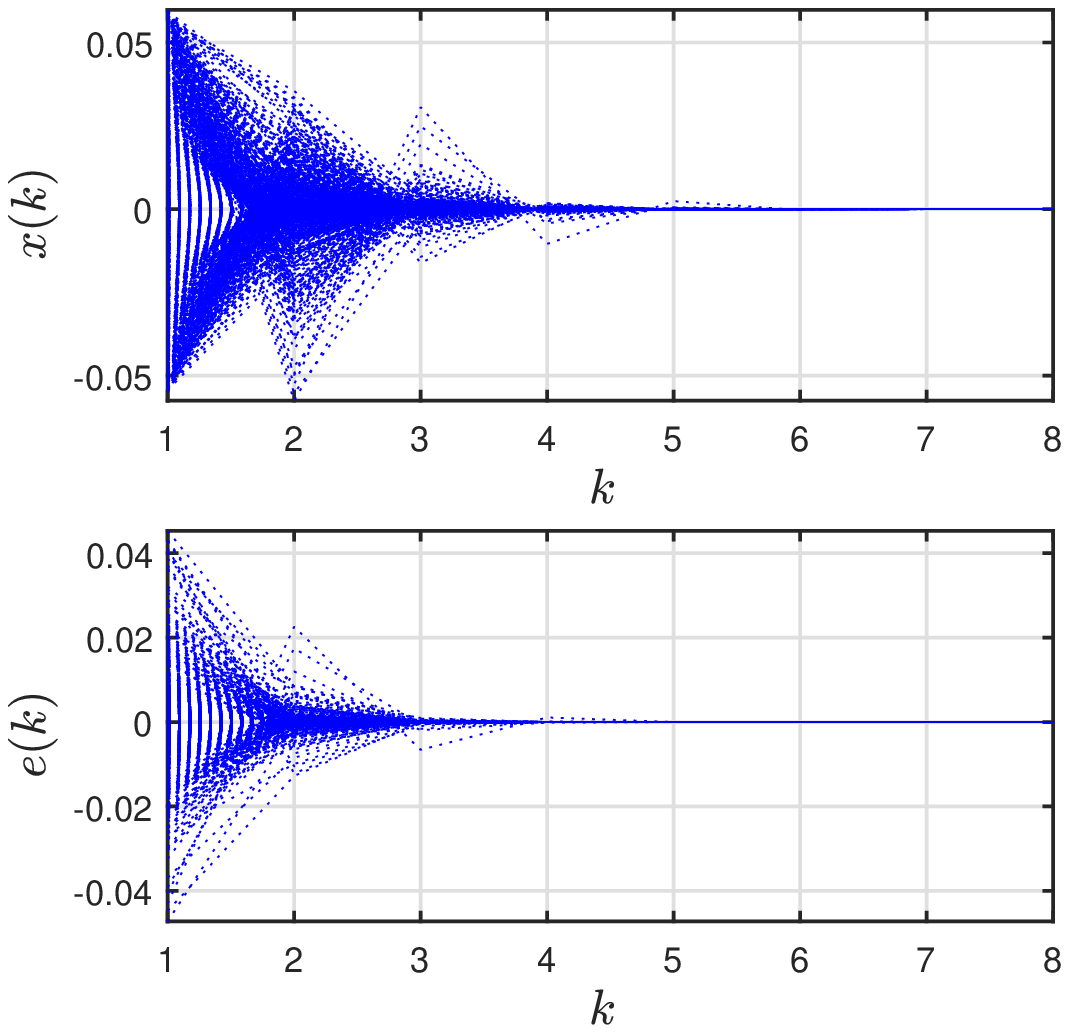}\\
		\parbox[c]{0.23\textwidth}{\footnotesize \centering (a)}
		\parbox[c]{0.23\textwidth}{\footnotesize \centering (b)}
		\caption{(a) Estimates $\hat{\mathbb{W}}_{\mathfrak{F}} (L)$, $\hat{\mathbb{X}}_{\mathfrak{F}} (L)$ of robust NDDs, estimate $\mathbb{X}_{\text{ls}} (L,0.0036)$ of robust DOA for closed-loops, controller training data and robust controller $\mu$. (b) State trajectories of closed-loops and noise trajectories.}
		\label{fig:verifyMethod}
	\end{center}
\end{figure}

The method proposed in this section is verified in the following example.

\begin{exmp} \label{exmp:control}
	Consider $\hat{f}, \delta$ and $L$ in Example~\ref{exmp:illuminate}. The interested region $\mathbb{W} = [-0.3,0.3] \times [-0.3,0.3] \subset \mathbb{R}^2$ in the state-control space is partitioned into $9\times10^4$ cells of size $0.002 \times 0.002$. The number of data points in $W^d$ is selected as $10^6$. For each data point $(x^d;u^d)$ in $W^d$, the number of data points in $\bar{X}^d_{\mathfrak{F}} (x^d,u^d)$ is selected as $200$. Using Algorithm~\ref{alg:est_W_F(L)}, an estimate $\hat{\mathbb{W}}_{\mathfrak{F}}(L)$ of the RNDD-SC $\mathbb{W}_{\mathfrak{F}}(L)$ is obtained and shown in Figure~\ref{fig:verifyMethod} (a) denoted by gray region. An estimate $\hat{\mathbb{X}}_{\mathfrak{F}}(L)$ of the RNDD-S $\mathbb{X}_{\mathfrak{F}}(L)$ is also obtained and shown in Figure~\ref{fig:verifyMethod} (a) denoted by green line segment in $x$-axis, where the neighborhood $\mathbb{X}_0$ of the origin that is not contained by $\hat{\mathbb{X}}_{\mathfrak{F}}(L)$ is $[-0.004, 0.004]$.  
	
	By solving the optimization problem \eqref{eq:optimization_alpha}, we obtain the largest level-set $\mathbb{X}_{\text{ls}}(L,0.0036) = [-0.06,0.06] \subset \mathbb{R}$ of $L(x) = x^2$ as the estimate of the RDOA for closed-loops, which is shown in Figure~\ref{fig:verifyMethod} (a) denoted by the blue line segment in $x$-axis. In order to find a controller $\mu$ belonging to the gray region, we select a training data set shown by red 'x' in Figure~\ref{fig:verifyMethod} (a). A robust controller $\mu$ is obtained using Gaussian processes regression, as shown in Figure~\ref{fig:verifyMethod} (a) denoted by black line.
	
	To verify whether the controller $\mu$ can stabilize all plants in the plant set for all initial state in $\mathbb{X}_{\text{ls}}(L,0.0036)$, we consider the controlled system $x(k+1) = \hat{f}(x(k),u(k)) + e(k)$, where noise $e(k)$ is drawn from the uniform distribution on $[-\delta(k),\delta(k)] \subset \mathbb{R}$ and $\delta(k) = \delta(x(k),u(k))$. Figure~\ref{fig:verifyMethod} (b) shows 1000 state trajectories of $x(k+1) = \hat{f}(x(k),\mu(x(k))) + e(k)$, whose initial states are drawn from the uniform distribution on $\mathbb{X}_{\text{ls}}(L,0.0036)$. We see that all state trajectories converge to the origin. Figure~\ref{fig:verifyMethod} (b) also shows 1000 noises trajectories corresponding to the 1000 state trajectories.
\end{exmp}

\section{Conclusion} \label{sec:conclusion}

In order to overcome the drawback of existing nonlinear robust control approaches, this paper proposes a new robust control method where the uncertain system is described by a non-affine nonlinear plant set. Under this circumstance, it is in general hard to fulfill the global stabilization, which requests an extensive investigation about the robust DOA of closed-loops. To this end, the sufficient condition is presented for robust asymptotic stabilization of the plant set and estimation of the RDOA for closed-loops. Moreover, due to non-affine nonlinearities, it is hard to obtain analytic solutions of the RNDDs. To overcome this problem, a data-driven method of estimating the RNDDs is proposed.


\bibliographystyle{plain}        
\bibliography{dd_robust_stab}       

\end{document}